\documentclass{article}
\usepackage{srcltx}
\usepackage{amsmath}
\usepackage{amssymb}
\usepackage{amsfonts}
\usepackage{latexsym}
\usepackage{textcomp}
\usepackage{appendix}
\usepackage{multirow}
\usepackage{booktabs}
\usepackage{subfigure}
\usepackage{epsfig}
\usepackage{url}
\usepackage{array}
\usepackage{marvosym}
\usepackage{graphics}
\usepackage{graphicx}

\title{Efficiency and credit ratings: a permutation-information-theory analysis}

\author{Aurelio Fern\'andez Bariviera\\ \scriptsize{Department of Business, Universitat Rovira i Virgili, Av. Universitat 1, 43204 Reus, Spain} \\ \scriptsize{\ttfamily aurelio.fernandez@urv.net} \and Luciano Zunino \\ \scriptsize{Centro de Investigaciones \'Opticas (CONICET La Plata - CIC),}\\ \scriptsize{C.C. 3, 1897 Gonnet, Argentina} \\ \scriptsize{Departamento de Ciencias B\'asicas, Facultad de Ingenier\'ia,}\\ \scriptsize{Universidad Nacional de La Plata (UNLP), 1900 La Plata, Argentina} \\ \scriptsize \ttfamily{lucianoz@ciop.unlp.edu.ar} \and M. Bel\'en Guercio \\  \scriptsize{Instituto de Investigaciones Econ\'omicas y Sociales del Sur, UNS-CONICET.} \\  \scriptsize{12 de Octubre y San Juan, B8000CTX Bah\'{\i}a Blanca, Argentina.} \\ \scriptsize{Universidad Provincial del  Sudoeste (UPSO).} \\ \scriptsize{ Alvarado 328, B8000CJH Bah\'ia Blanca, Argentina} \and Lisana B. Martinez \\   \scriptsize{Instituto de Investigaciones Econ\'omicas y Sociales del Sur, UNS-CONICET.} \\  \scriptsize{12 de Octubre y San Juan, B8000CTX Bah\'{\i}a Blanca, Argentina.} \\ \scriptsize{Universidad Provincial del  Sudoeste (UPSO).} \\ \scriptsize{ Alvarado 328, B8000CJH Bah\'ia Blanca, Argentina} \and Osvaldo A. Rosso \\ \scriptsize{Instituto de F\'{\i}sica, Universidade Federal de Alagoas (UFAL). } \\ 
\scriptsize{BR 104 Norte km 97, 57072-970 Macei\'o, Alagoas, Brazil. }\\ \scriptsize{Instituto Tecnol\'ogico de Buenos Aires (ITBA),} \\ \scriptsize{Av. Eduardo Madero 399, C1106ACD Ciudad Aut\'onoma de Buenos Aires, Argentina.}}

\begin{document}
\maketitle

\begin{abstract}
The role of credit rating agencies has been under severe scrutiny after the subprime crisis. In this paper we explore the relationship between credit ratings and informational efficiency of a sample of thirty nine corporate bonds of US oil and energy companies from April 2008 to November 2012. For that purpose, we use a powerful statistical tool relatively new in the financial literature: the \textit{complexity-entropy causality plane}. This representation space allows to graphically classify the different bonds according to their degree of informational efficiency. We find that this classification agrees with the credit ratings assigned by Moody's. Particularly, we detect the formation of two clusters, that correspond to the global categories of investment and speculative grades. Regarding to the latter cluster, two subgroups reflect distinct levels of efficiency. Additionally, we also find an intriguing absence of correlation between informational efficiency and firm characteristics. This allows us to conclude that the proposed permutation-information-theory approach provides an alternative practical way to justify bond classification. 

\end{abstract}

\section{Introduction}
\label{sec:int}
In his classical definition, \cite{Fama70} establishes that a market is informationally efficient if prices reflect all available information and classifies efficiency into three broad categories: (i) \textit{weak efficiency} if today price reflects the information embedded in the series of past prices, (ii) \textit{semi-strong efficiency} if prices reflect all public information, beyond past prices, and (iii) \textit{strong efficiency} if prices reflect all public and private information. We will center our study in the weak form of the informational efficiency. Prices are in fact a mechanism of signaling. \cite{Hayek45} says that the price system can be regarded as a platform for communicating information and its functioning is based on the economy of knowledge where ``by a kind of symbol, only the most essential information is passed on, and passed on only to those concerned''. In fact such a definition was anticipated in \cite{Gibson1889}, who wrote ``when shares become publicly known in an open market, the value which they acquire there may be regarded as the judgment of the best intelligence concerning them''. Later, \cite{Bachelier} formalized the first mathematical model of security prices, considering an stochastic process without memory. In fact, as recognized by \cite{LeRoy89}, the Efficient Market Hypothesis (EMH) is the theory of competitive equilibrium applied to securities markets. There is a vast literature on empirical research related to weak informational efficiency. It is worth mentioning here that deviations from the EMH have been confirmed for oil and energy markets - see, for instance, results obtained by \cite{paper:serletis2007}, \cite{paper:uritskaya2008}, \cite{Alvarez-Ramirez08}, \cite{paper:charles2009}, \cite{Alvarez-Ramirez2010}, \cite{Alvarez-Ramirez2010b} and \cite{paper:arouri2010}.

Since the subprime crisis credit rating agencies (CRAs) activities are under scrutiny, due to their difficulty to rank financial securities, specially Collateralized Debt Obligations (CDOs). There are some paradigmatic examples of slow reaction of CRAs to market movements: Enron was rated investment grade by both, Moody's and Standard \& Poor's, four days prior to its bankruptcy on December 2nd 2001, and more recently, Lehman Brother was still rated investment grade by both agencies on the day of its bankruptcy filing on September 15th 2008.

The aim of this paper is to build a bridge between credit ratings and a market derived measure, i.e. informational efficiency. Specifically, we explore the link between the correlated stochastic behavior of the bond yield time series (quantified by a combination of entropy and complexity measures) and bond ratings. We want to: (i) classify corporate bonds by means of the complexity-entropy causality plane, (ii) establish a correspondence between credit ratings and levels of informational efficiency, and (iii) analyze the potential link between market efficiency and firm ratios. It is worth remarking that we are not trying to analyze causality between credit rating and informational efficiency. Probably, a better rating induces market participants to trade more actively this bond, increasing the informative flow to the market and, consequently, the associated informational efficiency. On the contrary, it could be thought that better informational efficiency reflects an intrinsic quality of the firm that is captured by CRAs.

This paper is organized as follows. Section~\ref{sec:credit_retings} presents a review about credit ratings. Section~\ref{sec:inf_eff} describes the study of informational efficiency by using tools derived from Information Theory. Section~\ref{sec:data_results} details the data analyzed and the results obtained. Section~\ref{sec:conclusions} draws the main findings of our work. Finally, \ref{sec:Quantifiers} includes a thorough explanation of the permutation-information-theory based quantifiers applied in this paper.

\section{Credit ratings}
\label{sec:credit_retings}
Credit rating industry was born at the beginning of the twentieth century. In fact, Moody's, the oldest of the rating agencies, begun operations in 1909 publishing ratings of railroads companies. Credit ratings are intended to measure the likelihood of a firm to fulfill its debt obligations. According to the \cite{EC1060} ``credit rating means an opinion regarding the creditworthiness of an entity, a debt or financial obligation, debt security, preferred share or other financial instrument, or of an issuer of such a debt or financial obligation, debt security, preferred share or other financial instrument, issued using an established and defined ranking system of rating categories''. Nevertheless, credit ratings should not be understood as a buy or sell recommendation or a warranty of payment.

Credit ratings are based on the opinions of analysts who, after examining quantitative (e.g. financial statements, financial ratios, macroeconomic data, etc.) and qualitative data (interview to the firm's senior managers), reach a decision about the rating. These ratings are usually materialized in a letter grade or a combination of letters and figures. The best rating is Aaa and the worst is C. Between them, there are twenty categories, which intend to fine tune the credit appraisal. They can be divided into two global categories: investment (the upper half) and speculative (the bottom half) grades. The first category represents instruments issued by companies that have solid financial indicators and have a high payment capacity. The second category represents instruments issued by companies whose financial strength is questioned, and default on debt or late repayment is likely to occur.
 
At the beginning credit ratings were mostly focused on utility companies. Later, financial regulations increased the importance of CRAs. The Federal Reserve was created in 1913 and the Securities and Exchange Commission (SEC) of the United States in 1934. These two institutions rapidly begun to issue financial regulations to standardize the format of financial information disclosure and to limit the risk exposure of banks. In order to promote a safer and more transparent financial system, bank regulators in many countries link the minimum capital requirements to bond rating, as an standard measure of the riskiness of the bonds. Although there are a lot of CRAs, three of them, namely Moody's, Standard \& Poor's and Fitch, concentrate about 80\% of the market share. Successive regulations in the US and other countries favor this concentration and, as \cite{White2010} recognizes, ``creditworthiness judgments of these third-party raters attained the force of law''. In this line, \cite{paper:tichy2011} states that these three big rating agencies are regarded as all-powerful and unregulated elements of the financial markets.

A main criticism of the CRAs assessment is their lack of transparency. They do not disclose any of the criteria on which their ratings are based on and the methods applied. In addition, their business model is questioned. At the beginning of their activities, CRAs raised money from investors who were interested in an independent credit risk appraisal. However, in the 1970s, the revenue source shifted from `investor pays' to `issuer pays'. This change casts some doubts about the independency of opinions and, consequently, gives rise to potential conflict of interests. CRAs face two important forces: on the one hand they are interested in giving their best opinion on the credit quality of a firm, but, on the other hand, they want to give `good news' to the firm that pays for the service. This situation is relevant because in most markets it is mandatory for a firm to contract one of the CRAs in order to be authorized to issue new debts. Nevertheless, the fact that investors pay for rating information does not prevent from big mistakes. \cite{Ross1976} remembers that 11\% of investment grade corporate issues and 78\% of the municipal bonds that had been rated as Aaa or Aa defaulted during the Great Depression. This situation reflects the fact that credit rating is not a trivial task.

Most of the analyses on ratings in the literature focused their attention on guessing the variables that mainly affect or determine the assignation of rating categories, and on investigating the influence of rating changes in prices and yields. There are several papers that analyze the relationship between rating and firm risk, i.e. the comparison between ratings and market based measures to assess the risk of firms. In other words, they try to figure out if ratings inject new information to the market or, on the contrary, the market has already incorporated that information before the disclosure of ratings. \cite{Weinstein77} finds empirical evidence that prices change before rating change announcements. In the same line, \cite{GeskeDelianedis} conclude that markets anticipate to rating changes independently that the change is within or between speculative and investment grade categories. \cite{EltonGruber2004} analyze the corporate bond price behaviors, taking into account that the credit rating qualifications given by Moody's and Standard \& Poor's determine homogeneous groups related to the yield curve. However, they recognize the existence of other important factors that affect the price bond evolution such as the default risk, liquidity, tax liability, recovery rate and maturity. \cite{Perraudin04} find ``that bond prices and credit ratings generally embed similar views on relative credit risk, and any difference in views that do arise are often temporary''. In a recent work, \cite{May2010} detects that changes in credit ratings produce a significant reaction in bond prices. \cite{LuChenLiao2010} study the effect of uncertainty and asymmetry of information over corporate bond spreads and find that yield spread of bonds with short maturities are partially explained by corporate credit rating. \cite{Abadetal12} find evidence that negative announcements of ratings convey relevant information to the market.

In this paper we are particularly interested to explore about a potential relationship between credit ratings and informational efficiency. This link could help to find a more objective and unbiased way to classify credit ratings.

\section{Informational efficiency and the complexity-entropy causality plane}
\label{sec:inf_eff}
The study of weak informational efficiency is about the possibility of unveiling information from prices or return time series. Financial markets can be regarded as dynamical systems, whose behavior is recorded in time series of prices, yields, turnover, etc. These time series should be carefully analyzed in order to understand the underlying phenomenon. Quantifiers derived from Information Theory can be suitable candidates for this task since they allow to extract some properties of the probability distributions estimated from the observed data. One of the key metrics is Shannon entropy. Given any arbitrary discrete probability distribution $P=\{p_i \geq 0, i=1,\dots,M\}$ for which $\sum_{i=1}^{M} p_i=1$, Shannon entropy is defined as $S[P]=-\sum_{i=1}^M p_i \ln p_i$ \cite{book:cover2006}. It is equal to zero if the underlying structure is fully deterministic ($p_k=1 \wedge p_i=0,~\forall i \neq k$) and reaches a maximum value for an uncorrelated stochastic process (uniform distribution, i.e. $p_i = 1 / M,~\forall i = 1, \cdots, M$). It is remarkable that this amount of information is computed in terms of state probabilities and does not depend on a particular distribution.

The use of informational entropy to study economic phenomena can be traced back to \cite{TheilLeenders65}, where entropy is used to predict short-term price fluctuations in the Amsterdam Stock Exchange. \cite{Fama65entropy} and \cite{Dryden68} perform a similar study for the New York and London Stock Exchanges, respectively. \cite{PhilippatosNawrocki73} analyze the proportions of securities with positive, negative and null returns on the American Stock Exchange using Information Theory methodologies and conclude that they are dependent on the previous day and not significantly influenced by the proportion of untraded securities. \cite{PhilippatosWilson74} propose the average mutual information or shared entropy as a proxy for systematic risk. Much more recently, \cite{Risso08} combines entropy quantifier and symbolic time series analysis in order to relate informational efficiency and the probability of having an economic crash. Later, \cite{Risso09} also uses Shannon entropy to rank the informational efficiency of several stock markets around the world. Alvarez-Ramirez and coworkers implement a multiscale entropy concept to monitor the evolution of the informational efficiency over different time horizons. By applying this methodology to the crude oil prices \cite{Martina2011,OrtizCruz12} and the Dow Jones Index (DJI) \cite{paper:alvarez2012}, the presence of some particular cyclic dynamics is unveiled.

However, analyzing time series by only estimating the Shannon entropy could be insufficient since, as recalled by \cite{FeldmanCrutchfield98}, an entropy measure does not quantify the degree of structure presents in a process. In fact, it is necessary to measure the statistical complexity in order to fully characterize the system's dynamics. This is why we have proposed to consider also the statistical complexity for the analysis of financial time series \cite{paper:zunino2010}. Measures of statistical complexities try to quantify the underlying hidden organizational structure. In that sense, perfect order and maximal randomness (a periodic sequence and a fair coin toss, for example) are defined with zero complexity because they are the easiest to describe and understand. At a given distance from these extremes, a wide range of possible degrees of physical structure exists. The complexity measure allows to quantify this array of behavior \cite{FeldmanMcTague08}.

In the present paper we employ the \textit{complexity-entropy causality plane} (CECP), i.e. the representation space with the \textit{permutation entropy} of the system in the horizontal axis and an appropriate \textit{permutation statistical complexity} measure in the vertical one, for the analysis. This novel information-theory-tool was recently shown to be a practical and robust way to discriminate the linear and nonlinear correlations present in stock \cite{paper:zunino2010}, commodity \cite{paper:zunino2011} and sovereign bond \cite{Zunino2012} markets. The location in the CECP allows to quantify the inefficiency of the system under analysis because the presence of temporal patterns derives in deviations from the ideal position associated with a totally random process, i.e. normalized entropy and statistical complexity equal to one and zero, respectively. Consequently, the distance to this random ideal planar location can be used to define a ranking of efficiency. Technical details about the estimation of the permutation entropy, permutation statistical complexity as well as the construction of the CECP are left to~\ref{sec:Quantifiers} in order to make easy the reading of the paper. Readers unfamiliar with these topics are strongly encouraged to read it at this point.

\section{Data and results}
\label{sec:data_results}
In this work we analyze the daily values of the yield to maturity of thirty nine corporate bonds, corresponding to oil and energy companies of the United States. The yield to maturity is the annualized percentage return of a bond held until its stated maturity. All data were collected from Datastream database. The period under analysis goes from 1st April 2008 until 16th November 2012, giving a total of $N=1209$ data points for each bond. Time counting was performed over trading days, skipping weekends and holidays. In order to homogenize our sample, bonds are selected using the following criteria: (i) they are issued by companies from the United States and from the oil and energy sectors, (ii) they are straight bonds, (iii) the coupons are constant with a fixed frequency, (iv) the repayment is at par, (v) they mature before year 2029, (vi) they are long term (10 to 30 years), and (vii) the rating is available. To arrive to the thirty nine corporate bonds, we select only one bond by firm in order to avoid over representation bias. Codes and names of these indices are listed in Table~\ref{data}.

\begin{table}
\begin{scriptsize}
\caption{Corporate bonds.}
\begin{center}
\begin{tabular}{l l l l l l l}
\hline 
& DS Code & Borrower & Issue & Maturity & Coupon & Moody's \\
& & & date & date & & rating \\
\hline \\
1. & 18797N & Anadarko Petroleum Corporation & 15/10/96 & 15/10/26 & 7.500 & Ba1 \\
2. & 17874U & Apache Corporation & 27/02/96 & 15/03/26 & 7.700 & A3 \\
3. & 81359Q & Berry Petroleum Co. & 24/10/06 & 01/11/16 & 8.250 & B3 \\
4. & 56161F & Buckeye Partners Lp & 30/06/05 & 01/07/17 & 5.125 & Baa3 \\
5. & 90269K & Canadian Natural Resources Limited & 19/03/07 & 15/05/17 & 5.700 & Baa1 \\
6. & 93510X & Cimarex Energy Company & 01/05/07 & 01/05/17 & 7.125 & Ba2 \\
7. & 216442 & Conocophillips Company & 01/07/97 & 01/01/27 & 7.800 & A1 \\
8. & 18739L & El Paso Corporation & 03/02/98 & 01/02/18 & 7.000 & Ba3 \\
9. & 18258R & El Paso Natural Gas Company & 13/11/96 & 15/11/26 & 7.500 & Baa3 \\
10. & 18483P & Enbridge Energy Partners Lp & 01/10/98 & 01/10/28 & 7.125 & Baa1 \\
11. & 61384H & Energen Corporation & 22/07/97 & 24/07/17 & 7.400 & Baa3 \\
12. & 81359L & Energy Transfer Partners Ltd. & 23/10/06 & 15/02/17 & 6.125 & Baa3 \\
13. & 1688NK & EOG Resources Incorporated & 10/09/07 & 15/09/17 & 5.875 & A3 \\
14. & 2101ME & Forest Oil Corp. & 15/12/07 & 15/06/19 & 7.250 & B1 \\
15. & 251985 & Hess Corporation & 01/10/99 & 01/10/29 & 7.875 & Baa2 \\
16. & 46082X & Husky Energy Incorporated & 18/06/04 & 15/06/19 & 6.150 & Baa2 \\
17. & 18791U & Marathon Oil Corporation & 15/05/92 & 15/05/22 & 9.375 & Baa2 \\
18. & 1798EF & Mcmoran Exploration Co. & 14/11/07 & 15/11/14 & 11.875 & Caa1 \\
19. & 241094 & Murphy Oil Corporation & 04/05/99 & 01/05/29 & 7.050 & Baa3 \\
20. & 95576F & Nexen Incorporated & 04/05/07 & 15/05/17 & 5.650 & Baa3 \\
21. & 18616C & Noble Energy Incorporated & 15/10/93 & 15/10/23 & 7.250 & Baa2 \\
22. & 251451 & Occidental Petroleum Corporation & 10/02/99 & 15/02/29 & 8.450 & A2 \\
23. & 38803V & Overseas Shipholding Group Inc & 19/02/04 & 15/02/24 & 7.500 & Caa1 \\
24. & 80814U & Peabody Energy Corp & 12/10/06 & 01/11/26 & 7.875 & Ba1 \\
25. & 65954C & Pioneer Natural Resources Company & 01/05/06 & 01/05/18 & 6.875 & Ba1 \\
26. & 49481N & Plains All American Pipeline Lp & 15/02/05 & 15/08/16 & 5.875 & Baa3 \\
27. & 86422F & Plains Exploration \& Production Co. & 13/03/07 & 15/03/17 & 7.000 & B1 \\
28. & 64349H & Quicksilver Resources Incorporated & 16/03/06 & 01/04/16 & 7.125 & B3 \\
29. & 1722EQ & Range Resources Corporation & 28/09/07 & 01/10/17 & 7.500 & Ba3 \\
30. & 18737W & Sherwin-Williams Company & 10/02/97 & 01/02/27 & 7.375 & A3 \\
31. & 18242J & Spectra Energy Capital Llc & 20/07/98 & 15/07/18 & 6.750 & Baa2 \\
32. & 49480M & Stone Energy Corp. & 15/12/04 & 15/12/14 & 6.750 & Caa2 \\
33. & 18725W & Sunoco Incorporated & 01/11/94 & 01/11/24 & 9.000 & Ba2 \\
34. & 96571M & Swift Energy Company & 01/06/07 & 01/06/17 & 7.125 & B3 \\
35. & 602040 & Talisman Energy Incorporated & 21/10/97 & 15/10/27 & 7.250 & Baa2 \\
36. & 244814 & Tennessee Gas Pipeline Company & 13/03/97 & 01/04/17 & 7.500 & Baa3 \\
37. & 16527D & Transcanada Pipelines & 14/10/97 & 14/10/25 & 7.060 & A3 \\
38. & 243459 & Transcontinental Gas Pipe Line Co. & 15/07/96 & 15/07/26 & 7.080 & Baa2 \\
39. & 18241H & Valero Energy Corporation & 25/06/96 & 01/07/26 & 7.650 & Baa2 \\
\hline
\end{tabular}
\label{data}
\end{center}
\end{scriptsize}
\end{table}

\begin{figure}
\begin{center}
\subfigure{
\resizebox{0.75\columnwidth}{!}{\includegraphics{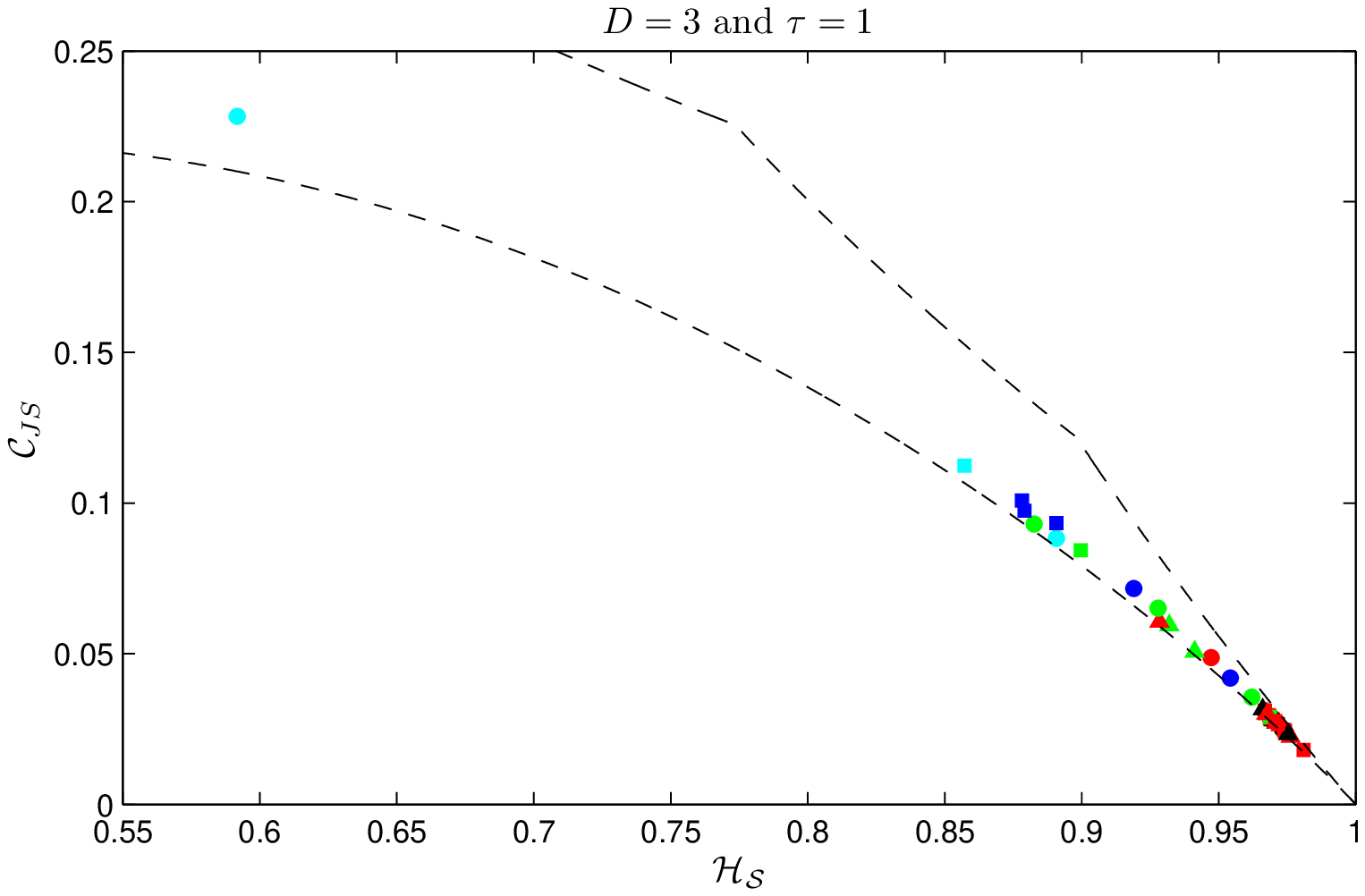}}}
\subfigure{
\resizebox{0.75\columnwidth}{!}{\includegraphics{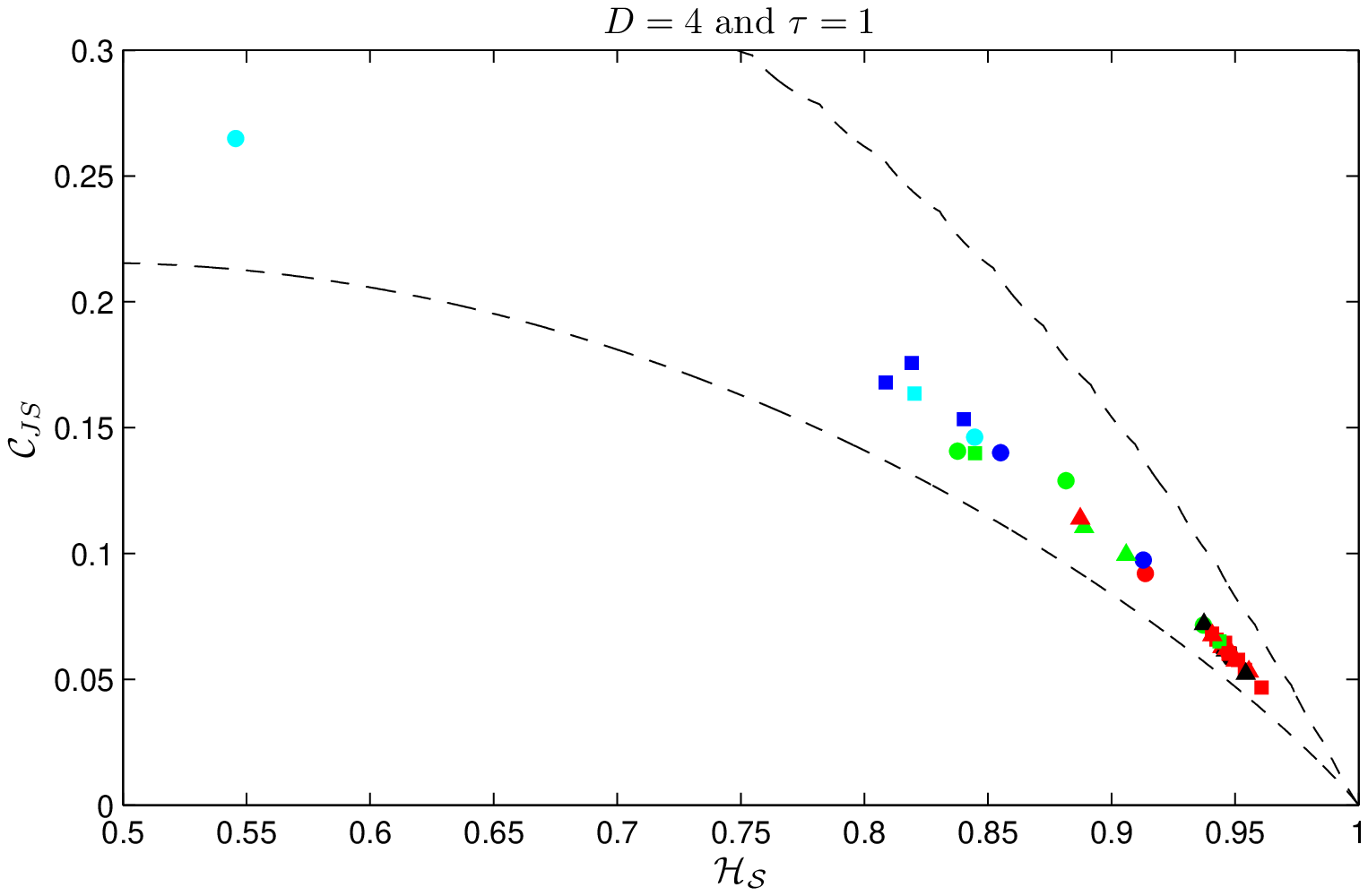}}}
\subfigure{
\resizebox{0.75\columnwidth}{!}{\includegraphics{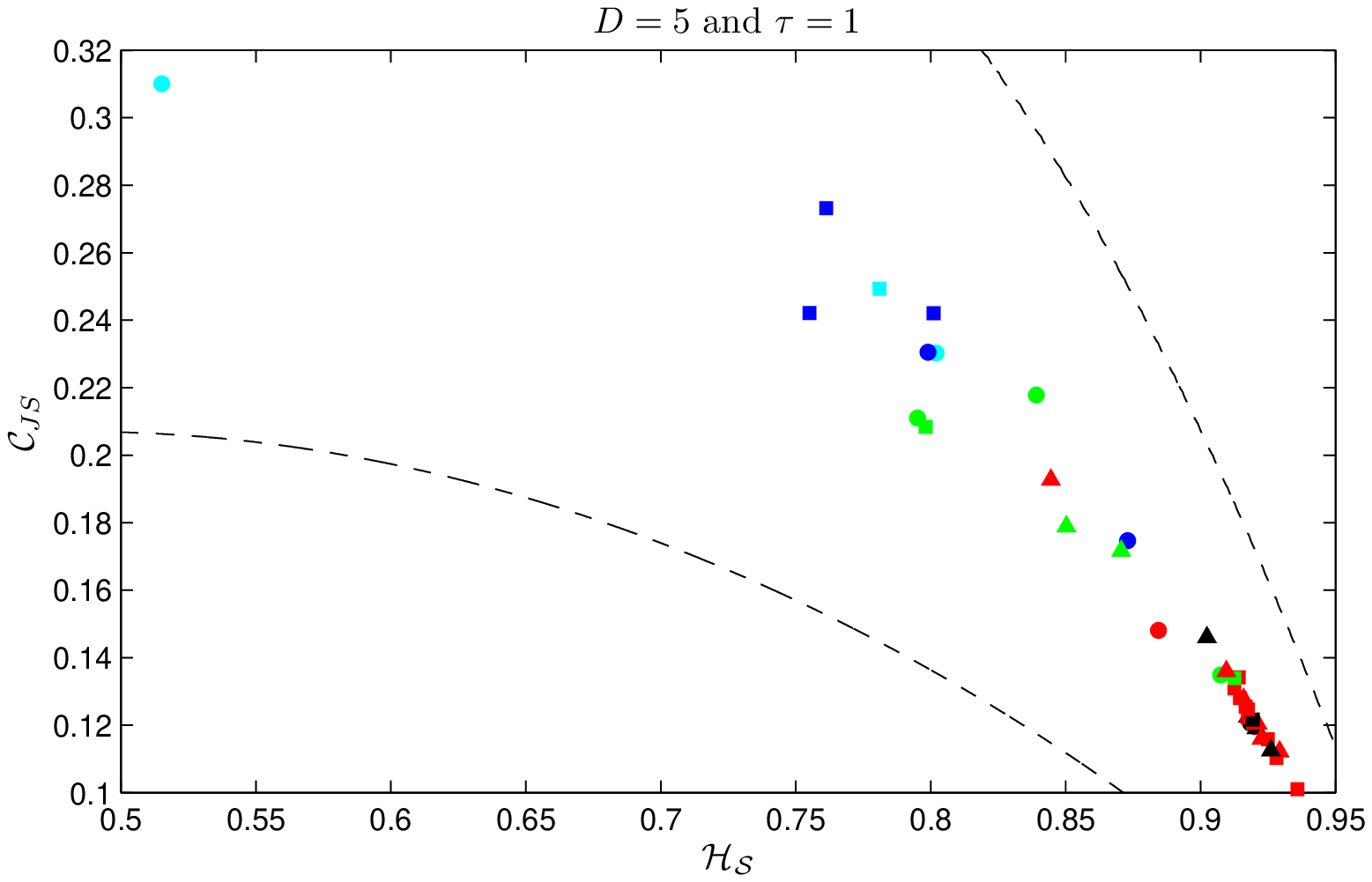}}}
\caption{(Color online) Location of the corporate bond markets of US oil and energy companies in the CECP. Permutation quantifiers are estimated by using different embedding dimensions $D=\{3,4,5\}$ and embedding delay $\tau=1$. The following symbols and colors are employed to discriminate the different credit ratings: black circle (A1), black square (A2), black triangle (A3), red circle (Baa1), red square (Baa2), red triangle (Baa3), green circle (Ba1), green square (Ba2), green triangle (Ba3), blue circle (B1), blue square (B3), light blue circle (Caa1) and light blue square (Caa2). Dashed lines represent the maximum and minimum complexity values for a given value of the entropy (see, for instance, \cite{paper:martin2006} for further details about these bounds).}
\label{cecp}
\end{center}
\end{figure}

\begin{figure}
\begin{center}
\subfigure{
\resizebox{0.75\columnwidth}{!}{\includegraphics{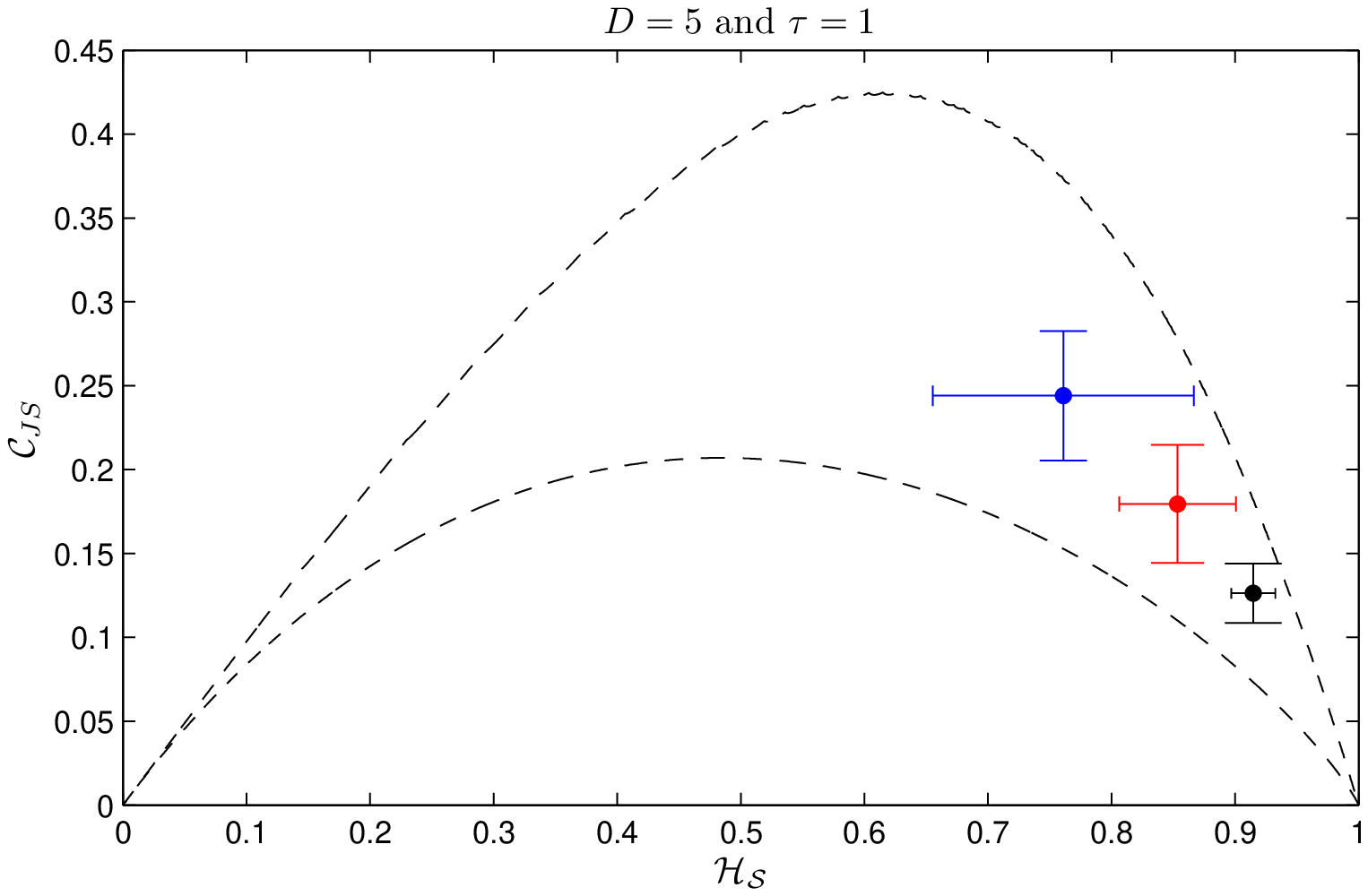}}}
\caption{Average location of the three clusters identified in the analysis: corporate bonds in the investment category grade (A1 to Baa3, black lines), in the upper part of the speculative grade (Ba1 to Ba3, red lines), and in the lower part of this category (B1 to Caa2, blue lines). Mean and standard deviation of the permutation quantifiers estimated with $D=5$ and embedding delay $\tau=1$ for each cluster are plotted. Dashed lines represent the maximum and minimum complexity values for a given value of the entropy.}
\label{average}
\end{center}
\end{figure}

We estimate permutation entropy ($\mathcal{H_S}$) and permutation statistical complexity ($\mathcal{C}_{JS}$) for the different corporate bonds. Embedding dimensions $D=\{3,4,5\}$ and embedding delay $\tau=1$ are chosen. Ordinal patterns generated by these parameters correspond to three, four and five consecutive days. It should be noted that long-range correlations in the time series are reflected in the relative frequency of these ordinal patterns, i.e. some particular patterns appear more often than the others due to the memory effect, and the estimated probability distribution results different from the uniform one expected for an uncorrelated stochastic process. Moreover, stochastic processes with different long-range correlations, such as fractional Gaussian noise (fGn) and fractional Brownian motion (fBm), can be clearly discriminated by using permutation quantifiers with parameters similar to that employed in the present analysis \cite{paper:bandt2007,paper:rosso2007a,paper:rosso2007d,paper:zunino2008}. Results obtained by estimating both permutation quantifiers, $\mathcal{H_S}$ and $\mathcal{C}_{JS}$, for the corporate bond markets of US oil and energy companies are displayed in Fig.~\ref{cecp}. According to them we can clearly identify two clusters. The planar location of each cluster corresponds to homogeneous rating categories: investment and speculative grades. The first category (black and red symbols) exhibits, in average, high entropy and low complexity indicating that series behaves more randomly and, thus, that they are closer to the informational efficient behavior. The second category can be subdivided into two subgroups. The bonds in the upper part of this category (Ba1 to Ba3), represented by green symbols, are located in an area with intermediate entropy and complexity values. Finally, the lower part of the speculative grade bonds (B1 to Caa2), indicated by blue and light blue symbols, exhibits less entropy and higher complexity, which highlight their informational inefficiency. Of course, these are average behaviors, and particular exceptions can be observed. In Fig.~\ref{average} the mean and standard deviation of the permutation quantifier values estimated with $D=5$ and embedding delay $\tau=1$ for the above mentioned clusters are displayed. Estimated values for permutation entropy and permutation statistical complexity show a high correlation. This kind of information redundancy is characteristic of stochastic processes. When the temporal correlations in the process increase, $\mathcal{H_S}$ and $\mathcal{C}_{JS}$ are smaller and larger, respectively, advising about the presence of these patterns. The locations in the CECP allow to conclude that the correlated stochastic properties of the yields time series is coherent with classifications given by CRAs. Following the same approach, \cite{Zunino2012} find that the qualifications given by Moody's to sovereign bonds of thirty countries are coherent with the location of the associated time series in the CECP.

\begin{figure}
\begin{center}
\subfigure{
\resizebox{0.75\columnwidth}{!}{\includegraphics{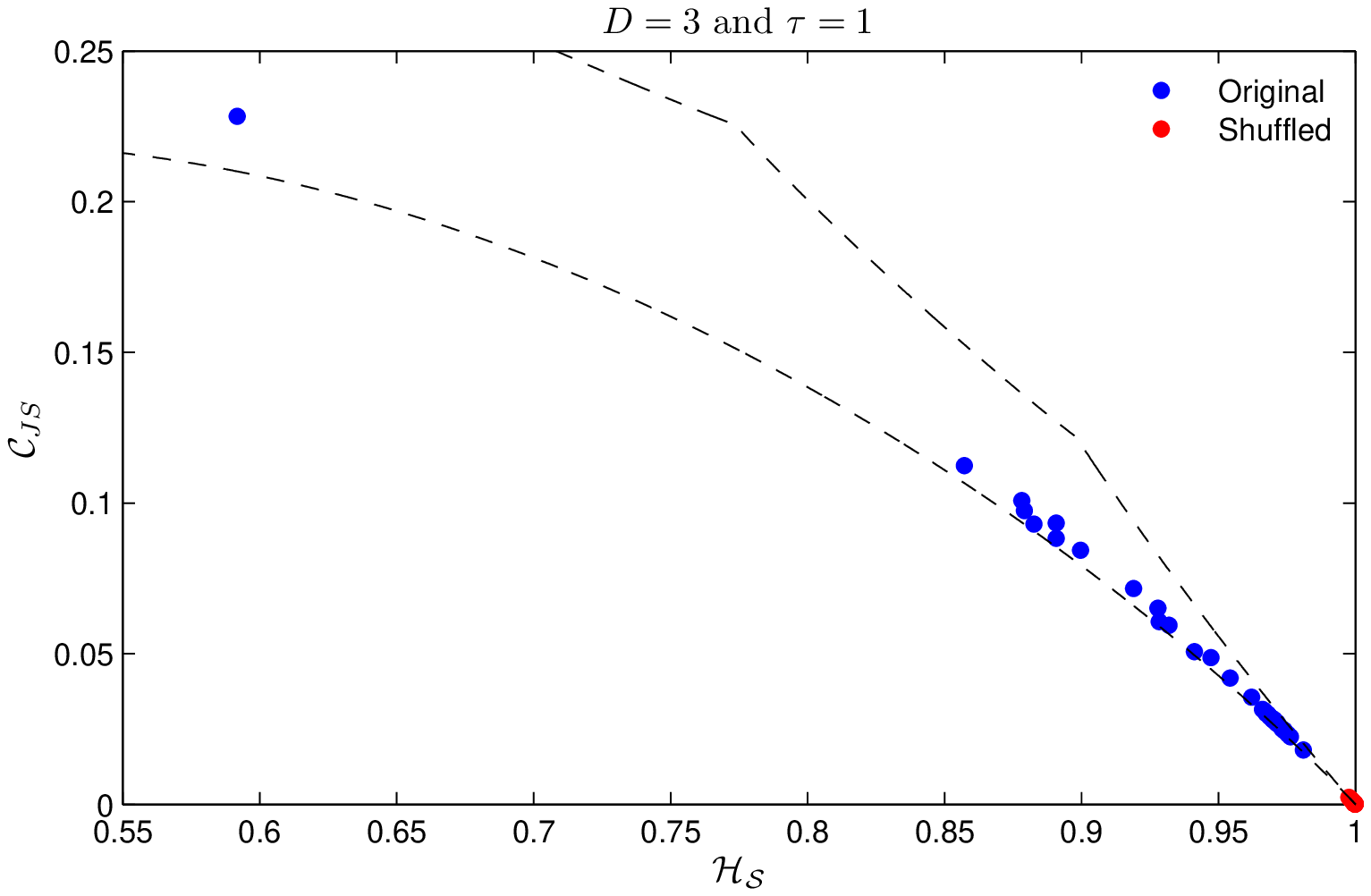}}}
\subfigure{
\resizebox{0.75\columnwidth}{!}{\includegraphics{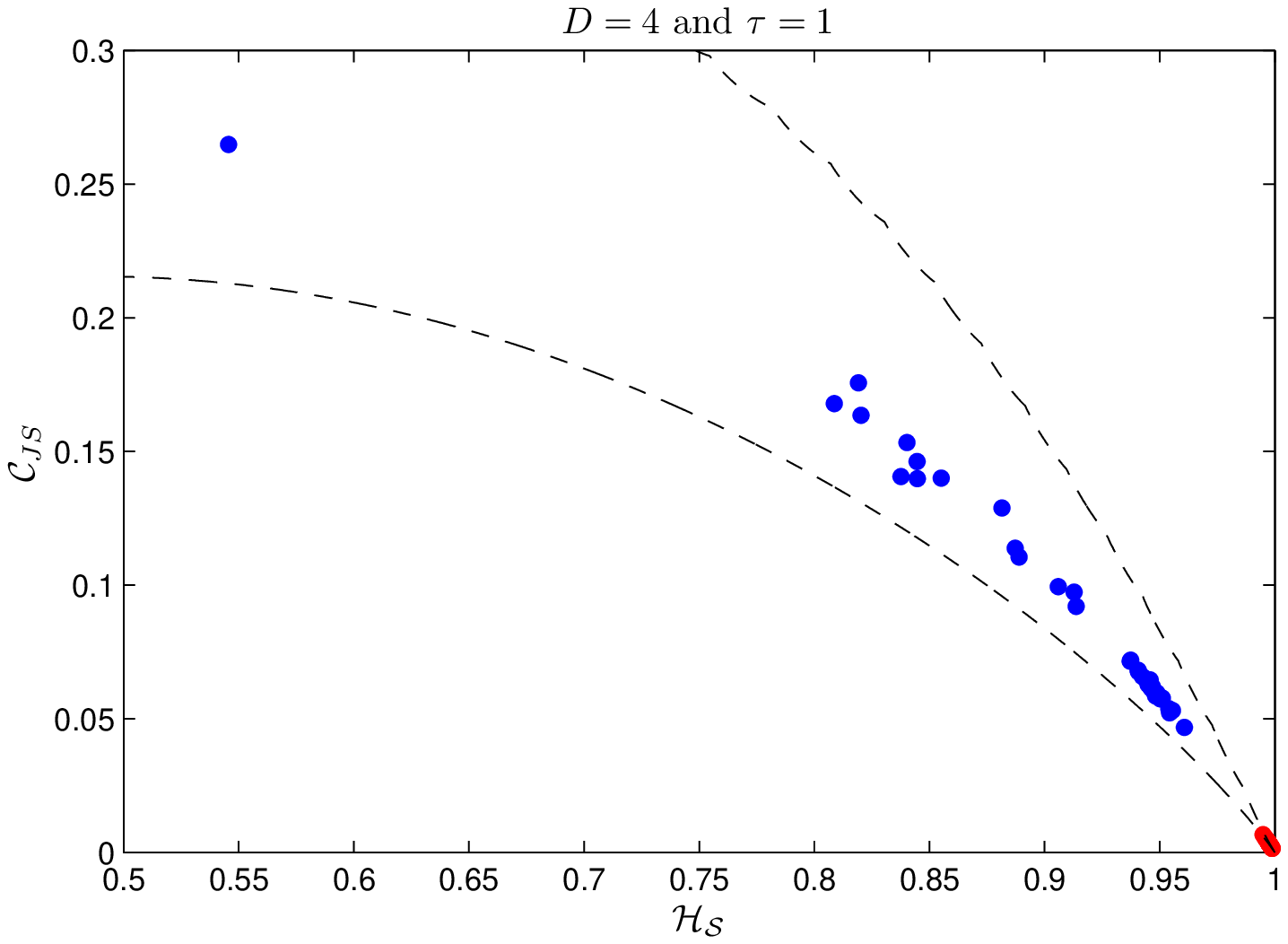}}}
\subfigure{
\resizebox{0.75\columnwidth}{!}{\includegraphics{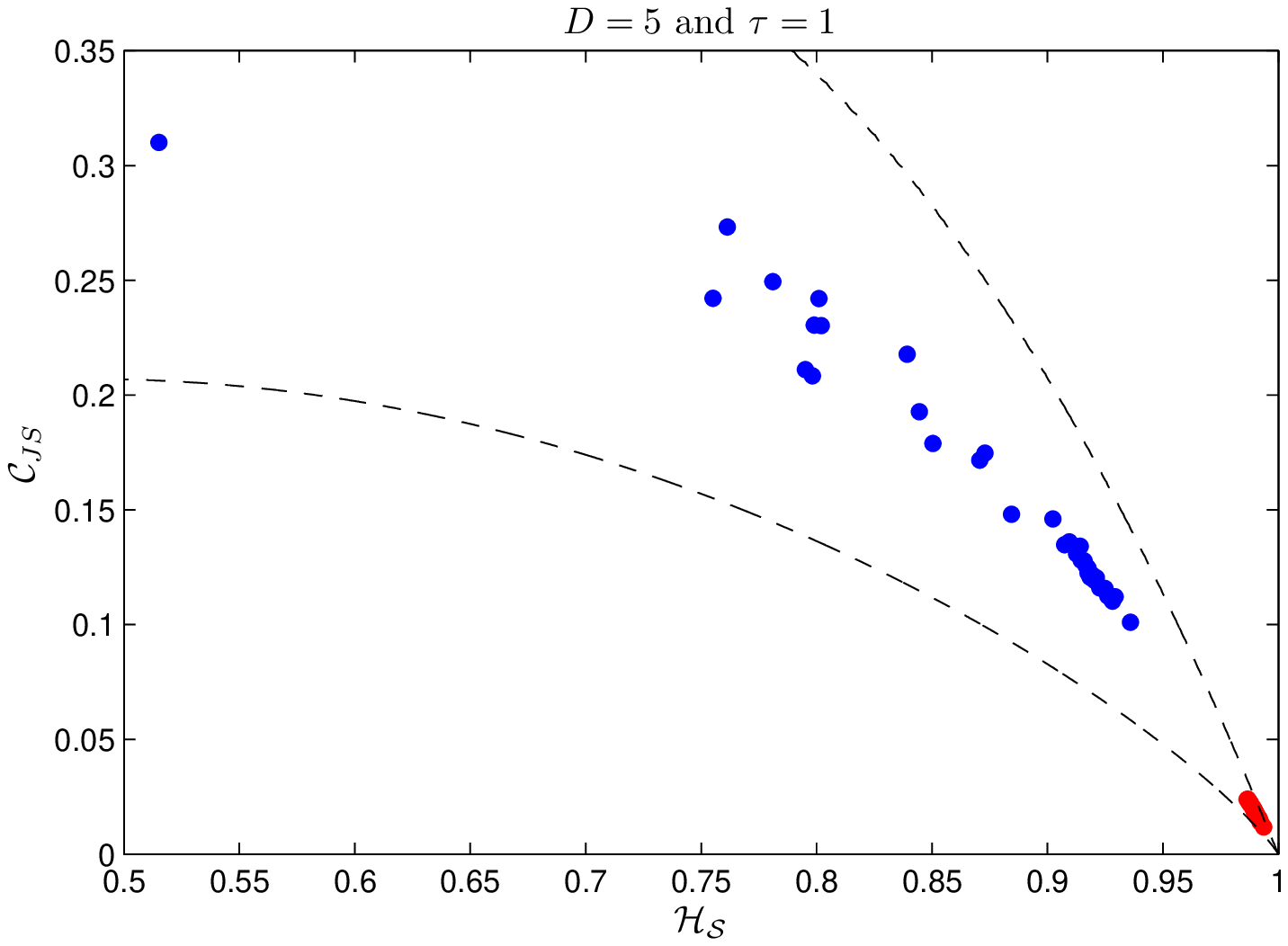}}}
\caption{(Color online) Location of the original (blue circles) and shuffled (red circles) corporate bond markets in the CECP with embedding dimensions $D=3$ (upper plot), $D=4$ (central plot) and $D=5$ (lower plot) and embedding delay $\tau=1$. The estimations for the shuffled realization are very close to that expected for a fully random record with $\mathcal{H_S} \approx 1$ and $\mathcal{C}_{JS} \approx 0$. Dashed lines represent the maximum and minimum complexity values for a given value of the entropy.}
\label{shuffled}
\end{center}
\end{figure}

Trying to justify the fact that intrinsic temporal correlations play a significant role in the CECP location obtained for the different corporate bonds, we have also estimated the permutation quantifiers for the shuffled corporate bond data. Shuffled realization of the original data are obtained permuting them in a random order, and eliminating, consequently, all non-trivial temporal correlations. From Fig.~\ref{shuffled}, where the location obtained for the original data (blue circles) and its shuffled counterparts (red circles) are depicted, it can be easily concluded that the positions obtained from original data are not obtained by chance and the underlying correlations are relevant. Estimated values of the permutation quantifiers for the shuffled data are very close to that expected for a fully random record ($\mathcal{H_S} \approx 1$ and $\mathcal{C}_{JS} \approx 0$). Stock \cite{paper:zunino2010b} and commodity \cite{paper:zunino2011} markets show a similar behavior.

In order to analyze the relationship between informational efficiency and firm characteristics, we perform Spearman's non-parametric correlation between permutation entropy and the average of some accounting ratios for the period 2008-2011. Selected accounting ratios and their expected signs are detailed in Table~\ref{ratiossign}. Results for this non-parametric correlation are shown in Table~\ref{Spearman}. Similar results were also obtained using Kendall's non-parametric correlation. If we consider the whole sample, there is not a statistical significant correlation. We have also computed the correlations for investment and speculative grade subgroups looking for a differential effect within bond categories. After this classification, the results are similar, but the correlation with the interest cash is significant, albeit with the opposite expected sign.

\begin{table}
\begin{scriptsize}
\caption{Description of the five selected accounting ratios.}
\begin{center}
\begin{tabular}{l c}
\hline 
Ratio & Expected sign \\
$\textrm{Quick Ratio (QR)}= \frac{\textrm{Cash \& short term investments}}{\textrm{Current liabilities}}$ & + \\
$\textrm{Current Ratio (CurrRatio)}= \frac{\textrm{Current assets}}{\textrm{Current liabilities}}$ & + \\
$\textrm{Coverage Ratio (CovRatio)}= \frac{\textrm{Gross income}}{\textrm{Interest expense on debt}}$ & + \\
$\textrm{Interest Earning ratio (IE)}= \frac{\textrm{Interest expense on debt}}{\textrm{Earnings before interest \& taxes}}$ & - \\
$\textrm{Interest Cash ratio (IC)}= \frac{\textrm{Interest expense on debt}}{\textrm{Net cash flow-operating activities}}$ & - \\
\hline
\end{tabular}
\label{ratiossign}
\end{center}
\end{scriptsize}
\end{table}

\begin{table}
\begin{scriptsize}
\caption{Spearman's non-parametric correlation between permutation entropy estimations and the five selected accounting ratios for the different embedding dimensions $D=\{3,4,5\}$. QR: Quick Ratio, CurrRatio: Current Ratio, CovRatio: Coverage Ratio, IE: Interest Earning ratio \& IC: Interest Cash ratio.}
\begin{center}
\begin{tabular}{rrrrrrrr}
\hline
& & & \multicolumn{1}{c}{QR} & \multicolumn{1}{c}{CurrRatio} & \multicolumn{1}{c}{CovRatio} & \multicolumn{1}{c}{IE} & \multicolumn{1}{c}{IC} \\ \hline
\multicolumn{1}{r}{\multirow{7}[2]{*}{All firms}} & \multicolumn{1}{r}{\multirow{2}[1]{*}{$D=3$}} & \multicolumn{1}{l}{Spearman's Rho} & -0.190 & -0.085 & 0.218 & 0.129 & 0.027 \\
\multicolumn{1}{r}{} & \multicolumn{1}{r}{} & \multicolumn{1}{l}{p-value} & 0.260 & 0.607 & 0.182 & 0.435 & 0.873 \\
\cline{2-8}
\multicolumn{1}{r}{} & \multicolumn{1}{r}{\multirow{2}[0]{*}{$D=4$}} & \multicolumn{1}{l}{Spearman's Rho} & -0.228 & -0.062 & 0.231 & 0.141 & 0.055 \\
\multicolumn{1}{r}{} & \multicolumn{1}{r}{} & \multicolumn{1}{l}{p-value} & 0.175 & 0.706 & 0.157 & 0.392 & 0.740 \\
\cline{2-8}
\multicolumn{1}{r}{} & \multicolumn{1}{r}{\multirow{2}[0]{*}{$D=5$}} & \multicolumn{1}{l}{Spearman's Rho} & -0.181 & -0.117 & 0.151 & 0.209 & 0.116 \\
\multicolumn{1}{r}{} & \multicolumn{1}{r}{} & \multicolumn{1}{l}{p-value} & 0.283 & 0.477 & 0.358 & 0.202 & 0.480 \\
\cline{2-8}
& \multicolumn{1}{c}{} & \multicolumn{1}{l}{N} & 37\dag & 39 & 39 & 39 & 39 \\ \hline
\multicolumn{1}{r}{\multirow{7}[2]{*}{Investment grade}} & \multicolumn{1}{r}{\multirow{2}[1]{*}{$D=3$}} & \multicolumn{1}{l}{Spearman's Rho} & 0.334 & 0.167 & 0.003 & -0.125 & 0.053 \\
\multicolumn{1}{r}{} & \multicolumn{1}{r}{} & \multicolumn{1}{l}{p-value} & 0.129 & 0.436 & 0.987 & 0.560 & 0.806 \\
\cline{2-8}
\multicolumn{1}{r}{} & \multicolumn{1}{r}{\multirow{2}[0]{*}{$D=4$}} & \multicolumn{1}{l}{Spearman's Rho} & 0.230 & 0.221 & 0.077 & -0.140 & 0.071 \\
\multicolumn{1}{r}{} & \multicolumn{1}{r}{} & \multicolumn{1}{l}{p-value} & 0.304 & 0.300 & 0.722 & 0.514 & 0.741 \\
\cline{2-8}
\multicolumn{1}{r}{} & \multicolumn{1}{r}{\multirow{3}[1]{*}{$D=5$}} & \multicolumn{1}{l}{Spearman's Rho} & 0.290 & 0.075 & -0.053 & 0.018 & 0.190 \\
\multicolumn{1}{r}{} & \multicolumn{1}{r}{} & \multicolumn{1}{l}{p-value} & 0.191 & 0.728 & 0.806 & 0.933 & 0.375 \\
\cline{2-8}
\multicolumn{1}{r}{} & \multicolumn{1}{r}{} & \multicolumn{1}{l}{N} & 22\dag & 24 & 24 & 24 & 24 \\ \hline
\multicolumn{1}{r}{\multirow{7}[2]{*}{Speculative grade}} & \multicolumn{1}{r}{\multirow{2}[1]{*}{$D=3$}} & \multicolumn{1}{l}{Spearman's Rho} & -0.118 & -0.196 & 0.046 & 0.371 & 0.582* \\
\multicolumn{1}{r}{} & \multicolumn{1}{r}{} & \multicolumn{1}{l}{p-value} & 0.676 & 0.483 & 0.869 & 0.173 & 0.023 \\
\cline{2-8}
\multicolumn{1}{r}{} & \multicolumn{1}{r}{\multirow{2}[0]{*}{$D=4$}} & \multicolumn{1}{l}{Spearman's Rho} & -0.132 & -0.182 & -0.025 & 0.461 & 0.654** \\
\multicolumn{1}{r}{} & \multicolumn{1}{r}{} & \multicolumn{1}{l}{p-value} & 0.639 & 0.516 & 0.930 & 0.084 & 0.008 \\
\cline{2-8}
\multicolumn{1}{r}{} & \multicolumn{1}{r}{\multirow{3}[1]{*}{$D=5$}} & \multicolumn{1}{l}{Spearman's Rho} & -0.068 & -0.189 & -0.104 & 0.489 & 0.732** \\
\multicolumn{1}{r}{} & \multicolumn{1}{r}{} & \multicolumn{1}{l}{p-value} & 0.810 & 0.499 & 0.713 & 0.064 & 0.002 \\
\cline{2-8}
\multicolumn{1}{r}{} & \multicolumn{1}{r}{} & \multicolumn{1}{l}{N} & 15 & 15 & 15 & 15 & 15 \\
\hline
\end{tabular}
\label{Spearman}
\end{center}
\vspace{-0.5cm}
\dag~~El Paso Natural Gas Company and Tennessee Gas Pipeline Company are not included because cash \& short term investments figures are not available for these firms in the Datastream database.\\
\textasteriskcentered~~significant at the 5\% level.\\
\textasteriskcentered\textasteriskcentered~significant at the 1\% level.
\end{scriptsize}
\end{table}

According to these results, we conclude that accounting ratios are not significantly correlated with permutation entropy. This situation is relevant considering that accounting ratios are used for computing credit ratings and, as the CECP analysis shows, credit ratings are associated with informational efficiency. Consequently, accounting ratios and permutation entropy are both related to credit ratings but they appear to be independent of each other. In light of these results, we can conclude that permutation quantifiers provide additional explicative power to justify bond classification.

\section{Conclusions}
\label{sec:conclusions}
This paper analyzes the relationship between informational efficiency and credit ratings in the corporate bond markets of US oil and energy companies. On the one hand, we find that bonds that are included in the investment grade category, i.e. those with the highest creditworthiness, are located in the region of the CECP that represents the most informational efficient behavior. On the other hand, bonds that belong to the speculative grade category exhibit less entropy and greater complexity, which indicates less efficiency. These different planar locations confirm a significant relationship between informational efficiency and credit rating of corporate bonds. 

Additionally, in this paper, we investigate a potential link between entropy and accounting ratios. The aim of this approach is to explore if firm characteristics are related to the permutation entropy, and thus, connected with the degree of informational efficiency. Surprisingly, only one of the five selected accounting ratios, the interest cash, is statistically significant for speculative grade bonds. However, this correlation does not present the expected sign. This situation highlights an absence of correlation between accounting ratios and informational efficiency. Since we have confirmed a correlated behavior between informational efficiency and the permutation quantifiers estimated in this paper, we conclude that the CECP provides an alternative and more transparent way to justify bond classification.

In the future, we propose to further study the reasons behind the lack of relationship between accounting ratios and permutation entropy, in spite of the fact that, on the one hand, credit ratings and permutation entropy, and, on the other hand, credit ratings and accounting ratios, are actually linked. Additionally, we would like to expand this study to other sectors in order to set a comparative study and verify that the CECP provides a good classification of the rating quality, independently of the firm sector.

\section*{Acknowledgements}
Luciano Zunino, M. Bel\'en Guercio and Osvaldo A. Rosso were supported by Consejo Nacional de Investigaciones Cient\'ificas y T\'ecnicas (CONICET), Argentina. Lisana B. Martinez acknowledges the support of a Ph.D. scholarship from Department of Business of Universitat Rovira i Virgili, Spain. Osvaldo A. Rosso acknowledges support as a CNPq fellow, Brazil.

\appendix

\section{Permutation-information-theory based quantifiers}
\label{sec:Quantifiers}

\subsection{Entropy and Statistical Complexity}
\label{sec:Quantifiers1}
The information content of a system is usually evaluated via a probability distribution function (PDF) describing the apportionment of some measurable or observable quantity, commonly a time series $\mathcal{S}(t)$. An information measure can be roughly defined as a quantity that
characterizes this given probability distribution. The Shannon entropy is very often used as the most ``natural" one (\cite{book:shannon1949}). Given any arbitrary discrete probability distribution $P=\{p_i,i=1,\dots,M\}$, with $M$ the number of degrees of freedom, Shannon's logarithmic information measure reads
\begin{equation}
\mathrm{S}[P]=-\sum_{i=1}^{M} p_i \ln p_i.
\label{Shannon}
\end{equation}
It can be regarded as a measure of the uncertainty associated with the physical process described by $P$. If $\mathrm{S}[P]=\mathrm{S}_{min}=0$ we are in position to predict with complete certainty which of the possible outcomes $i$, whose probabilities are given by $p_i$, will actually take place. Our knowledge of the underlying process described by the probability distribution is maximal in this instance. In contrast, our knowledge is minimal for the equiprobable distribution $P_e=\{p_i=1/M,i=1,\dots,M\}$ and, consequently, the uncertainty is maximal, $\mathrm{S}[P_e]=\mathrm{S}_{max}$.

It is widely known that an entropic measure does not quantify the degree of structure presents in a process \cite{FeldmanCrutchfield98}. Moreover, it was recently shown that measures of statistical or structural complexity are necessary for a better understanding of chaotic time series because they are able capture their organizational properties \cite{FeldmanMcTague08}. This specific kind of information is not revealed by randomness' measures. Rosso and coworkers introduced an effective \textit{statistical complexity measure} (SCM) that is able to detect essential details of the dynamics and differentiate different degrees of periodicity and chaos \cite{Lamberti2004119}. This specific SCM, that provides important additional information regarding the peculiarities of the underlying probability distribution, is defined via the product\footnote{This functional product form for the SCM is due to \cite{LMC95}.}
\begin{equation}
\mathcal{C}_{JS}[P]=\mathcal{Q}_J[P, P_e] \cdot \mathcal{H}_{S}[P]
\label{Complexity}
\end{equation}
of the \textit{normalized Shannon entropy}
\begin{equation}
\mathcal{H}_{S}[P]=\mathrm{S}[P]/\mathrm{S}_{max},
\label{Shannon-normalizada}
\end{equation}
with $\mathrm{S}_{max}=\mathrm{S}[P_e]=\ln M$, ($0 \leq \mathcal{H}_S \leq 1$) and $P_e$ the equiprobable distribution, and the so-called \textit{disequilibrium} $\mathcal{Q}_{J}$. This latter quantifier is defined in terms of the extensive (in the thermodynamical sense) \textit{Jensen-Shannon divergence} $\mathcal{J}[P, P_e]$ that links two PDFs. We have
\begin{equation}
\mathcal{Q}_{J} [P,P_e]=Q_0 \cdot \mathcal{J} [P, P_e],
\label{Disequilibrium}
\end{equation}
with
\begin{equation}
\mathcal{J} [P,P_e]=\mathrm{S} \left[(P + P_e)/2 \right] - \mathrm{S}[P]/2 - \mathrm{S}[P_e]/2.
\label{Jensen}
\end{equation}
$Q_0$ is a normalization constant, equal to the inverse of the maximum possible value of $\mathcal{J} [P,P_e]$. This value is obtained when one of the values of $P$, say $p_m$, is equal to one and the remaining $p_i$ values are equal to zero, i.e.,
\begin{equation}
Q_0=-2 \left\{ \left(\frac{M+1}{M}\right)\ln(M+1)-2 \ln(2M) + \ln M \right\}^{-1}.
\label{Q0}
\end{equation}
The Jensen-Shannon divergence, that quantifies the difference between two (or more) probability distributions, is especially useful to compare the symbol-composition of different sequences \cite{paper:grosse2002}. We stress the fact that the statistical complexity defined above is the product of two normalized entropies (the Shannon entropy and the Jensen-Shannon divergence), but it is a non-trivial function of the entropy because it depends on two different probabilities distributions, i.e., the one corresponding to the state of the system, $P$, and the equiprobable distribution, $P_e$, taken as reference state. Furthermore, it has been shown that for a given value of $\mathcal{H}_S$, the range of possible SCM values varies between a minimum $\mathcal{C}_{min}$ and a maximum $\mathcal{C}_{max}$ \cite{paper:martin2006}. Therefore, the evaluation of the complexity provides additional insight into the details of the system's probability distribution, which is not discriminated by randomness measures like the entropy. It can also help to uncover information related to the correlational structure between the components of the physical process under study \cite{paper:rosso2009,paper:rosso2009b}.

\subsection{Complexity-entropy plane}
\label{sec:Quantifiers2}
In statistical mechanics one is often interested in isolated systems characterized by an initial, arbitrary, and discrete probability distribution, and the main purpose is to describe its evolution towards equilibrium. At equilibrium, we can suppose, without loss of generality, that this state is given by the equiprobable distribution $P_e$. The temporal evolution of the statistical complexity measure (SCM) can be analyzed using a two-dimensional (2D) diagram of $\mathcal{C}_{JS}$ versus time $t$. However, the second law of thermodynamics states that, for isolated systems, entropy grows monotonically with time ($d\mathcal{H}_S/dt \geq 0$). This implies that $\mathcal{H}_S$ can be regarded as an arrow of time, so that an equivalent way to study the temporal evolution of the SCM is through the analysis of $\mathcal{C}_{JS}$ versus $\mathcal{H}_S$. The \textit{complexity-entropy plane} has been used to study changes in the dynamics of a system originated by modifications of some characteristic parameters (see, for instance, \cite{paper:rosso2009c}, \cite{paper:zunino2010b,paper:zunino2011,Zunino2012} and references therein).

\subsection{Estimation of the Probability Distribution Function}
\label{sec:Quantifiers3}
When using quantifiers based on Information Theory, such as $\mathcal{H}_S$ and $\mathcal{C}_{JS}$, a probability distribution associated with the time series under analysis should be provided beforehand. Many methods have been proposed for a proper estimation of it. We can mention: (i) frequency counting \cite{paper:rosso2009c}, (ii) procedures based on amplitude statistics \cite{paper:demicco2008}, (iii) binary symbolic dynamics \cite{paper:mischaikow1999}, (iv) Fourier analysis \cite{paper:powell1979}, and (v) wavelet transform \cite{paper:rosso2001}, among others. Their applicability depends on particular characteristics of the data, such as stationarity, time series length, variation of the parameters, level of noise contamination, etc. In all these cases the dynamics' global aspects can be somehow captured, but the different approaches are not equivalent in their ability to discern all the relevant physical details.

Methods for symbolic analysis of time series that discretize the raw series and transform it into a sequence of symbols constitute a powerful tool. They efficiently analyze nonlinear data and exhibit low sensitivity to noise \cite{paper:finn2003}. However, finding a meaningful symbolic representation of the original series can be a subtle task \cite{paper:bollt2000}. Different symbolic sequences may be assigned to a given time series \cite{paper:daw2003}. In this respect, an issue of some importance is that of ascertaining whether the temporal order in which the distinct time series values appear is considered or not. In the first case one says that \textit{causal information} has been taken into account. If one merely assigns a symbol $a$ of the finite alphabet $\mathfrak{A}$ to each value of the time series, the ensuing symbolic sequence can be regarded as a non-causal coarse-grained description of the time series under consideration. The PDF extracted from the time series will not have any causal information. The usual histogram technique corresponds to this kind of assignment. Causal information may be incorporated into the construction process that yields $P$ if one symbol of a finite alphabet $\mathfrak{A}$ is assigned instead to a (phase-space) trajectory's portion, i.e., we assign ``words" to each trajectory portion. The Bandt and Pompe (BP) methodology \cite{BandtPompe02} for extracting a PDF from a time series corresponds to the causal type of assignment, and the resulting probability distribution $P$ constitutes, thus, a causal coarse-grained description of the system. ``Partitions'' are devised by comparing the order of neighboring relative values rather than by apportioning amplitudes according to different levels. The appropriate symbol sequence arises naturally from the time series. No model-based assumptions are needed. 

Given a time series $\mathcal{S}(t)=\{x_t;t=1,\dots,N\}$, an embedding dimension $D>1$ $(D \in \mathbb{N})$, and an embedding delay $\tau$ $(\tau \in \mathbb{N})$, the BP-pattern of order $D$ generated by
\begin{equation}
\label{eq:vectores}
s \mapsto \left(x_{s},x_{s+\tau},\dots,x_{s+(D-2)\tau},x_{s+(D-1)\tau}\right),
\end{equation}
is considered. To each time $s$, BP assign a $D$-dimensional vector that results from the evaluation of the time series at times $s,s+\tau,\dots,s+(D-2)\tau,s+(D-1)\tau$. Clearly, the higher the value of $D$, the more information about the ``future'' is incorporated
into the ensuing vectors. By the ordinal pattern of order $D$ related to the time $s$, BP mean the permutation $\pi = (r_0~r_1~\dots~r_{D-2}~r_{D-1})$ of $(0~1~\dots~D-2~D-1)$ defined by
\begin{equation}
\label{eq:permuta}
x_{s+r_0\tau} \le x_{s+r_{1}\tau} \le \dots x_{s+r_{D-2}\tau} \le x_{s+r_{D-1}\tau}.
\end{equation}
In this way the vector defined by Eq.~(\ref{eq:vectores}) is converted into a definite symbol $\pi$. To get a unique result BP consider that $r_i<r_{i+1}$ if $x_{s+r_{i}\tau} = x_{s+r_{i+1}\tau}$. This is justified if the values of ${x_t}$ have a continuous distribution so that equal values are very unusual. For all the $D!$ possible orderings (permutations) $\pi_i$ when embedding dimension is $D$, their associated relative frequencies can be naturally computed according to the number of times this particular order sequence is found in the time series, divided by the total number of sequences. Thus, an ordinal pattern probability distribution $P=\{p(\pi_i),i=1,\dots,D!\}$ is derived from the time series.

In order to illustrate the BP recipe, consider a simple example: a time series with seven ($N=7$) values $x=\{4,7,9,10,6,11,3\}$ and we evaluate the BP-PDF for $D=3$ and $\tau=1$. Triplets $(4,7,9)$ and $(7,9,10)$ represent the permutation pattern $(012)$ since the values are in increasing order. On the other hand, $(9,10,6)$ and $(6,11,3)$ correspond to the permutation pattern $(201)$ since $x_{s+2} \le x_s \le x_{s+1}$, 
while $(10,6,11)$ has the permutation pattern $(102)$ with $x_{s+1} \le x_s \le x_{s+2}$. Then, the associated probabilities result: $p_{(012)}=p_{(201)}=2/5$; $p_{(102)}=1/5$; $p_{(021)}=p_{(120)}=p_{(210)}=0$.

\begin{figure}
\begin{center}
\subfigure{
\resizebox{0.6\columnwidth}{!}{\includegraphics{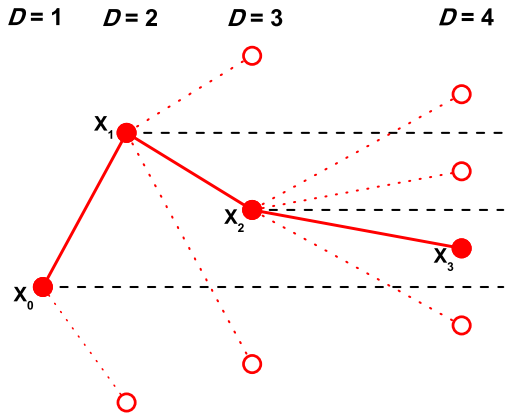}}}
\caption{Illustration of the construction-principle for ordinal patterns of length $D$ \cite{paper:parlitz2012}. For $D=4$, full circles and continuous lines represent the sequence $x_0 \le x_3 \le x_2 \le x_1$ which leads to the pattern $(0321)$.}
\label{fig:patrones}
\end{center}
\end{figure}

Graphically, Fig.~\ref{fig:patrones} illustrates the construction principle of the ordinal patterns of length $D=\{2,3,4\}$ \cite{paper:parlitz2012}. Consider the sequence $\{x_0,x_1,x_2,x_3\}$. For $D=2$, there are only two possible direction from $x_0$ to $x_1$, up and down. For $D=3$, starting from $x_1$ (up) the third part of the pattern can be above $x_1$, below $x_0$ or between $x_0$ and $x_1$ as it is illustrated in the Fig.~\ref{fig:patrones}. A similar situation is obtained starting from $x_1$ (down). For $D=4$, for each one of the 6 possible positions for $x_2$ we will have 4 possible locations for $x_3$, leading in this way finally to the $D!=4!=24$ different ordinal patterns. A graphical representation of all possible patterns corresponding to $D=\{3,4,5\}$ can be found in Fig.~2 of \cite{paper:parlitz2012}.

The BP-generated probability distribution $P$ is obtained once we fix the embedding dimension $D$ and the embedding delay $\tau$. The former parameter plays an important role in the evaluation of the appropriate probability distribution, since $D$ determines the number of accessible states, given by $D!$. Moreover, it has been established that the length $N$ of the time series must satisfy the condition $N \gg D!$ in order to achieve a reliable statistics and proper distinction between stochastic and deterministic dynamics \cite{paper:rosso2007a,paper:staniek2007}. With respect to the selection of the parameters, BP suggest in their cornerstone paper \cite{BandtPompe02} to work with $3 \leq D \leq 7$ with a time lag $\tau=1$. Nevertheless, other values of $\tau$ might provide additional information. \cite{paper:zunino2010} and \cite{paper:soriano2011b,paper:soriano2011a} have recently showed that this parameter is strongly related, when it is relevant, to the intrinsic time scales of the system under analysis.

It is clear that, applying this prescription for symbolizing time series, some details of the original amplitude information and variability are lost. However, a meaningful reduction of the complex systems to their basic inherent structure is provided. The symbolic representation of time series by recourse to a comparison of consecutive points ($\tau=1$) or non-consecutive ($\tau>1$) points allows for an accurate empirical reconstruction of the underlying phase-space, even in the presence of weak (observational and dynamical) noise \cite{BandtPompe02}. Furthermore, the ordinal pattern associated PDF is invariant with respect to nonlinear monotonous transformations. Accordingly, nonlinear drifts or scalings artificially introduced by a measurement device will not modify the quantifiers' estimation, a useful property if one deals with experimental data (see, i.e., \cite{paper:saco2010}). These advantages make the BP approach more convenient than conventional methods based on range partitioning. Additional advantages of the method reside in its simplicity (we need few parameters: the pattern length/embedding dimension $D$ and the embedding delay $\tau$) and the extremely fast nature of the pertinent calculation-process.

In this work the normalized Shannon entropy, $\mathcal{H}_S$ (Eq.~(\ref{Shannon-normalizada})), and the SCM, $\mathcal{C}_{JS}$ (Eq.~(\ref{Complexity})), are evaluated using the permutation probability distribution. Defined in this way, these quantifiers are usually known as \textit{permutation entropy} and \textit{permutation statistical complexity} \cite{paper:rosso2007d}. They characterize the diversity and correlational structure, respectively, of the orderings present in the complex time series. The \textit{complexity-entropy causality plane} (CECP) is defined as the two-dimensional (2D) diagram obtained by plotting permutation statistical complexity (vertical axis) versus permutation entropy (horizontal axis) for a given system \cite{paper:rosso2007a}. For further details about the estimation of permutation quantifiers and an exhaustive list of its main biomedical and econophysics applications we refer the readers to \cite{paper:zanin2012}.

\bibliographystyle{plain}  
\bibliography{EnergyMarket}

\end{document}